\documentclass[aps,pre,reprint,superscriptaddress,showpacs,preprintnumbers,amsmath,amssymb]{revtex4-2}
\usepackage{graphicx}
\usepackage{dcolumn}
\usepackage{bm}
\usepackage{color}
\begin{document}
\preprint{Preprint}
\title{Jamming transition and normal modes of polydispersed soft particle packing}
\author{Kuniyasu Saitoh}
\affiliation{Department of Physics, Faculty of Science, Kyoto Sangyo University, Kyoto 603-8555, Japan}
%
\author{Brian P. Tighe}
\affiliation{Delft University of Technology, Process \& Energy Laboratory, Leeghwaterstraat 39, 2628 CB Delft, The Netherlands}
\date{\today}
\begin{abstract}
The jamming transition of soft particles characterized by narrow size distributions has  been well studied by physicists.
However, polydispersed systems are more relevant to engineering, and the influence of polydispersity on jamming phenomena is still unexplored.
Here, we numerically investigate jamming transitions of polydispersed soft particles in two dimensions.
We find that polydispersity strongly influences contact forces, local coordination, and the jamming transition density. In contrast,
the critical scaling of pressure and elastic moduli is not affected by the particle size distribution. Consistent with this observation, we find that the vibrational density of states is also insensitive to the polydispersity.
Our results suggest that, regardless of particle size distributions,
both mechanical and vibrational properties of soft particle packings near jamming are governed by the distance to jamming.
\end{abstract}
\maketitle
%
\emph{Introduction} ---
Soft particles such as foams, emulsions, and granular materials are ubiquitous in our daily lives.
They are of great importance to technologies, including food, granular, and pharmaceutical products \cite{larson,review4}.
It is now well known that soft particles exhibit a rigidity transition,\
i.e.,\ the so-called \emph{jamming transition}, at critical packing fraction $\phi_c$ \cite{gn1,gn3,gn4}.
In recent years, critical behavior of their mechanical, geometrical, and rheological properties
(e.g., pressure, elastic moduli, excess coordination number, the first peak of radial distribution function, and viscosity) near jamming
has well been tested by numerous experiments and simulations \cite{gn2,rs0,gr1,rheol0,katgert13,rheol6}.
Furthermore, disordered configurations of jammed soft particles contrast sharply with periodic structures of regular lattices
such that their normal modes are distinct from those of usual solids \cite{ashcroft}.
For instance, the vibrational density of states (VDOS) of jammed soft particles exhibits a plateau extending down to zero frequency
as the system approaches the unjamming transition \cite{vm0,vm1,vm2,vm3,vm4,vm5}.
In addition, quasi-localized modes coexist with low-frequency vibrations \cite{qlv0,qlv1,qlv2,qlv3,qlv4,qlv5,qlv6,qlv7,qlv8,qlv9,qlv10,qlv11}
and the non-Debye scaling of VDOS is observed in between the low-frequency and plateau regimes \cite{nds0,nds1,nds2}.
It is also theoretically and numerically confirmed that the elastic \cite{aval_quasi0,aval_quasi1,saitoh13,CohesiveJamming1} and complex moduli \cite{rl0,rl3,saitoh16} are directly linked to the VDOS,
hence linear (visco)elasticity of soft particle packings can be predicted from knowledge of low frequency (long wavelength) vibrations of the particles \cite{saitoh13,CohesiveJamming1,rl0,rl3,saitoh16}.

Though the jamming transition and normal modes of soft particle packings have extensively been explored by the theories, experiments, and numerical simulations,
most previous works assumed that the particles are either monodispersed, bidispersed, or weakly polydispersed.
Since the seminal work by O'Hern et al. \cite{gn1} employed monodisperse systems in three dimensions and bidisperse mixtures of soft particles with a size ratio of $1.4$ in two dimensions, these systems have become canonical reference points. Much less attention has been paid to systems with broadly distributed particle sizes.
Nevertheless, polydisperse systems are intrinsically relevant to geophysics and civil engineering \cite{polydisperse2}, because grain sizes in a seismic fault are power law-distributed \cite{geology01,geology02}.
Moreover, the particle size distribution of Apollonian packings is given by a power law \cite{apollonian}.
There are also indications that polydispersity plays an important role in jamming and elasticity. Both the critical packing fraction $\phi_c$ and bulk modulus of bidisperse mixtures are sensitive to the size ratio \cite{bidisperse0}.
In addition, recent study of droplets with a power law distribution showed that pressure and $\phi_c$ are controlled by distribution's exponent \cite{polydisperse0}.
In contrast, the same study found that the distribution of coordination number is independent of the same exponent.
Similarly, an experimental study of polydispersed granular particles revealed that the macroscopic friction coefficient in the critical state is not affected by polydispersity \cite{polydisperse1}.
These contrasting results highlight the need for a systematic study of the interplay between polydispersity and its interplay with critical scaling near the jamming transition.

In this Communication, we report  results of simulations of polydispersed soft particles in $d = 2$ spatial dimensions generated with molecular dynamics (MD).
We analyze packings' statistics, geometry, and mechanics by systematically varying their packing fraction $\phi$ and their polydispersity,
quantified by a ratio between the maximum size to the minimum size, $\lambda$.
We find that some features are directly controlled by $\lambda$, and we quantify the form of this dependence.
These ``sensitive'' features include the distributions of local forces and coordination (which broaden dramatically) and the critical packing fraction (which increases).
Other features are surprisingly insensitive to $\lambda$. These include the mean coordination, pressure, elastic moduli, and VDOS.

\emph{Numerical methods} ---
We study two-dimensional polydispersed particles using MD simulations.
We model a repulsive force between the particles, $i$ and $j$, in contact by a linear spring as $\bm{f}_{ij} = k\delta_{ij}\bm{n}_{ij}$ with the stiffness $k$.
Here, $\bm{n}_{ij} = (\bm{r}_i-\bm{r}_j)/r_{ij}$ with the center-to-center distance, $r_{ij}\equiv|\bm{r}_i-\bm{r}_j|$, is a unit vector parallel to the normal direction,
where $\bm{r}_i$ ($\bm{r}_j$) is the position of the $i$-th ($j$-th) particle.
In addition, $\delta_{ij}\equiv d_{ij}-r_{ij}$ is introduced as the overlap between the particles, with $d_{ij}\equiv R_i+R_j$.
We randomly sample each particle radius $R_i$ from a power-law size distribution, $P(R_i)\propto R_i^{-\nu}$, with the exponent $\nu$.
The distribution function is limited to the range, $R_\mathrm{min}<R_i<R_\mathrm{max}$,
so that we can control polydispersity of the system by changing a \emph{size ratio}, $\lambda\equiv R_\mathrm{max}/R_\mathrm{min}$.
In this study, we fix $\nu=3$, which is typical of the size distributions of grains in seismic faults
\footnote{The power-law exponent for cumulative size distributions distributes around $2$ \cite{geology01,geology02} and thus $\nu=3$ is typical of the size distributions.},
and vary $\lambda$ from $2$ to $20$.
In Appendix \ref{sec:packing}, we show the size distributions $P(R_i)$.
We also examined the influence of the power-law exponent $\nu$ and confirmed that our results are qualitatively the same if the exponent is in the range, $3\leq\nu\leq 4$.

To make a static packing of polydispersed particles, we randomly distribute the $N=2048$ particles in a $L\times L$ square periodic box
such that packing fraction of the particles is given by $\phi=\sum_{i=1}^N\pi R_i^2/L^2$.
We then minimize the elastic energy $E=\sum_{i=1}^N\sum_{j>i} k\delta_{ij}^2/2$ using the FIRE algorithm \cite{FIRE} with all particle masses set to unity.

\begin{figure}
\includegraphics[width=\columnwidth]{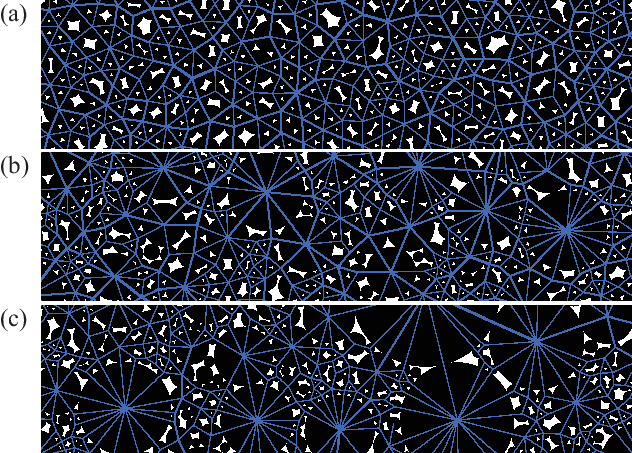}
\caption{
Snapshots of polydispersed particles (circles), where the packing fraction is $\phi=0.9$ and the size ratio increases as $\lambda=$ (a) $2$, (b) $10$, and (c) $20$.
The solid lines represent force-chains, where their width is proportional to the magnitude of repulsive force between the particles in contact,\ i.e.\ $k\delta_{ij}$.
\label{fig:fchain}}
\end{figure}

Figure \ref{fig:fchain} displays snapshots of static packings after minimization.
The packing fraction is given by $\phi=0.9$ and the size ratio increases from (a) $\lambda=2$ to (c) $20$.
See Appendix \ref{sec:packing} for a full image of (c).
If the size ratio is small, the force network (solid lines) is homogeneous in space
and the local coordination number for each particle varies little (Fig.\ \ref{fig:fchain}(a)).
However, as $\lambda$ increases, one observes that the forces become heterogeneous
and the coordination numbers for the largest particles are much larger than for smaller particles (Figs.\ \ref{fig:fchain}(b) and (c)).
%

\emph{Statistics of contact forces and coordination number} ---
The force heterogeneity evident in Fig.\ \ref{fig:fchain} is reflected in the distribution function of contact forces $P(f)$, which broadens with increasing polydispersity.
Figure \ref{fig:pdfz}(a) displays $P(f)$ for size ratios from $\lambda=2$ to $20$ (see Appendix \ref{sec:packing} for the full data set).
Here, $f$ represents the magnitude of repulsive force $\bm{f}_{ij}$,\ i.e.\ $k\delta_{ij}$, and the horizontal axis is scaled by the average $\langle f\rangle$ for each $\lambda$.
If $\lambda$ is small,\ e.g.\ $\lambda=2$, $P(f)$ is well fitted to a Gaussian distribution (solid line).
However, the tail broadens in highly polydispersed packings, such that $P(f)$ is better described with an exponential at large $f$, $P(f)\sim\exp\left(-f/\langle f\rangle\right)$ (dashed line).
A similar crossover from compressed-exponential to exponential tails has been observed as a result of other physical parameters, including increasing stress anisotropy \cite{en6,en7}, increasing spatial dimension \cite{gu5,en8}, increasing particle asphericity \cite{azema12}, and decreasing distance to the unjamming transition \cite{pdf_frl2}. 
Our results add polydispersity to this list.

The distribution of coordination number, $P(z)$, is also affected by the polydispersity.
Fig.~\ref{fig:pdfz}(b) shows that $P(z)$ broadens with increasing $\lambda$ (see Appendix \ref{sec:packing} for the full data set).
For sufficiently large  $\lambda$, the distribution approaches a power law, $P(z)\sim z^{-4.2}$ (dashed line), with a large-$z$ cutoff.
One might expect the cutoff to be proportional to the perimeter of the largest disks, and therefore linear in $\lambda$; instead we find $z^\ast \sim \lambda^{0.74}$
(see inset), which grows more slowly but still diverges.
These results complement a previous study \cite{polydisperse0}, which found power law decay for varying exponent $\nu$.
In addition, we note that the ``granocentric" model \cite{corwin_2009,corwin_2010} successfully reproduces the contact number distribution in narrowly polydisperse packings of emulsion droplets.
It may be possible to extend the model to broadly polydisperse packings; however the calculation is challenging and beyond the scope of this paper.

In Appendix \ref{sec:packing}, we also examine the radial distribution function, $g(r)$, of polydispersed packings.
We find that the first peak of $g(r)$ gets higher and both the first and second peaks shift to shorter distances with the increase of $\lambda$.
%
\begin{figure}
\includegraphics[width=\columnwidth]{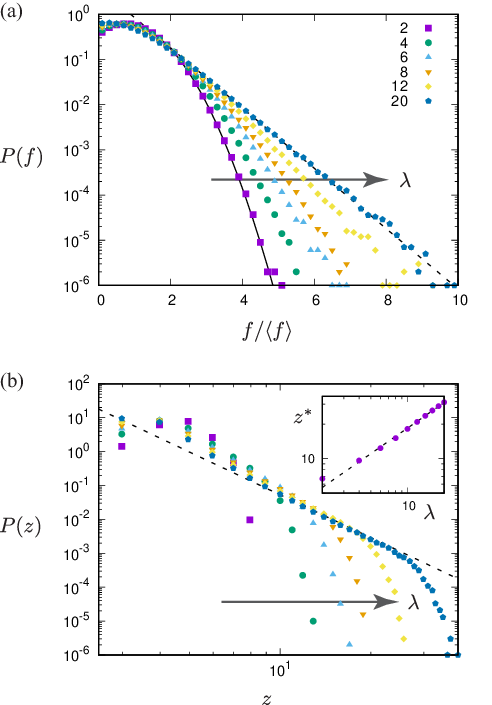}
\caption{
(a) Semi-logarithmic plots of the distribution function of contact forces $f$
and (b) double logarithmic plots of the distribution function of coordination number $z$.
The packing fraction of the particles is given by $\phi=0.90$ and the size ratio $\lambda$ increases as listed in the legend of (a) and indicated by the arrows.
In (a), the horizontal axis is scaled by the average $\langle f\rangle$ for each $\lambda$ and the solid line represents a Gaussian fit to the data of $\lambda=2$.
The dashed lines indicate (a) the exponential tail, $P(f)\sim\exp\left(-f/\langle f\rangle\right)$, and (b) power-law, $P(z)\sim z^{-4.2}$, for the data of $\lambda=20$.
The inset to (b) shows that the cutoff scales as $z^\ast\sim\lambda^{0.74}$ (dashed line).
\label{fig:pdfz}}
\end{figure}

\begin{figure}
\includegraphics[width=\columnwidth]{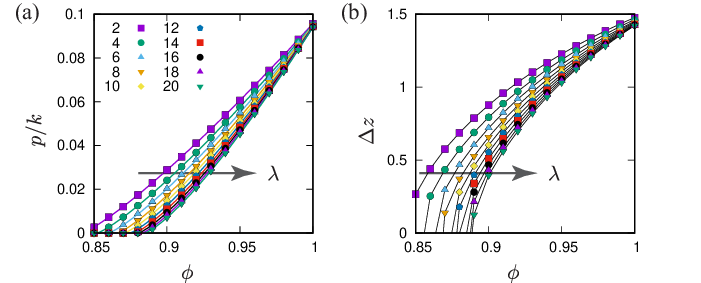}
\caption{
(a) The scaled pressure $p/k$ and (b) excess coordination number, $\Delta z\equiv\langle z\rangle-z_c$, as functions of the packing fraction $\phi$.
The size ratio $\lambda$ increases as listed in the legend of (a) and indicated by the arrows.
The solid lines in (a) represent fitting functions (see the text).
\label{fig:last}}
\end{figure}

\emph{Jamming transition and critical packing fraction} ---
We have shown that distribution functions are sensitive to polydispersity, as quantified by $\lambda$.
It is therefore natural to ask to what extent this polydispersity-dependence is inherited by macroscopic (averaged) quantities.
In canonical bidisperse packings ($\lambda = 1.4)$, the pressure $p$ and excess coordination number, $\Delta z \equiv \langle z\rangle-z_c$, scale as $p/k \sim (\phi-\phi_c)$ and $\Delta z \sim (\phi-\phi_c)^{1/2}$, respectively.
Here $\phi_c$ is  the critical packing fraction \cite{gn1,gn3,gn4}, $\langle z\rangle$ is the mean coordination number, and $z_c=2d-2d/N$ is the central force isostatic value (for $N$ particles in $d$-dimensions) \cite{finite0}.
We also calculate $p$ and $\Delta z$ of polydisperse packings to examine the effect of polydispersity on their scaling near jamming.
Figure \ref{fig:last} displays (a) the scaled pressure $p/k$ and (b) $\Delta z$ as functions of the packing fraction, where $\lambda$ increases as indicated by the arrows.
As can be seen, both $p/k$ and $\Delta z$ start to increase from zero at $\phi=\phi_c$, where $\phi_c$ shifts to higher values with the increase of $\lambda$.
When calculating $\Delta z$, we first remove mechanically unstable particles (``rattlers'') from the system.
In Appendix \ref{sec:packing}, we show that the fraction of rattlers linearly increases with $\lambda$ except for the case of $\lambda=1$ (monodisperse packings).

In Appendix \ref{sec:packing}, we also show the dependence of elastic energy $E$ and mean overlap $\langle\delta\rangle$ on the packing fraction.
As $p/k$ and $\Delta z$, both $E$ and $\langle\delta\rangle$ start to increase from zero at $\phi_c$, where $\phi_c$ increases with the increase of $\lambda$.
%

\begin{figure}
\includegraphics[width=\columnwidth]{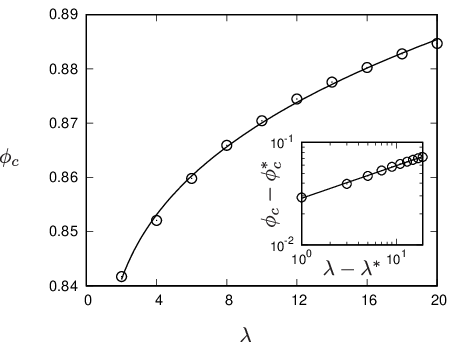}
\caption{
The critical packing fraction $\phi_c(\lambda)$ extracted from the data of $p/k$ (Fig.\ \ref{fig:last}(a)).
The inset is a double logarithmic plot of $\phi_c-\phi_c^\ast$ and $\lambda-\lambda^\ast$.
The solid lines represent the power-law, $\phi_c-\phi_c^\ast \sim \left(\lambda-\lambda^\ast\right)^{0.32}$ with $\lambda^\ast = 1$ and $\phi_c^\ast \simeq 0.81$.
\label{fig:jamming}}
\end{figure}

It is apparent from Fig.\ \ref{fig:last} that the jamming transition density is dependent on  polydispersity.
$\phi_c$ is higher in systems with higher polydispersity, because small particles can fill voids between larger ones.
To quantify the $\lambda$-dependence of $\phi_c$, we fit a power law to each $p/k$ dataset and extrapolate the $x$-intercept.
The results are shown in Figure \ref{fig:jamming}, and are well described by the power-law, $\phi_c-\phi_c^\ast \sim \left(\lambda-\lambda^\ast\right)^{0.32}$ (solid line).
Here, $\lambda^\ast = 1$ and $\phi_c^\ast \simeq 0.81$ indicate the size ratio and critical density for monodispersed particles, respectively.
While we have been unable to find a theoretical explanation for the specific value of the exponent,
we note that similar power-law shifts in the critical packing fraction also occur in packings of ellipsoidal particles \cite{donev04} and sticky particles \cite{CohesiveJamming0,CohesiveJamming1}.
%

\begin{figure}
\includegraphics[width=\columnwidth]{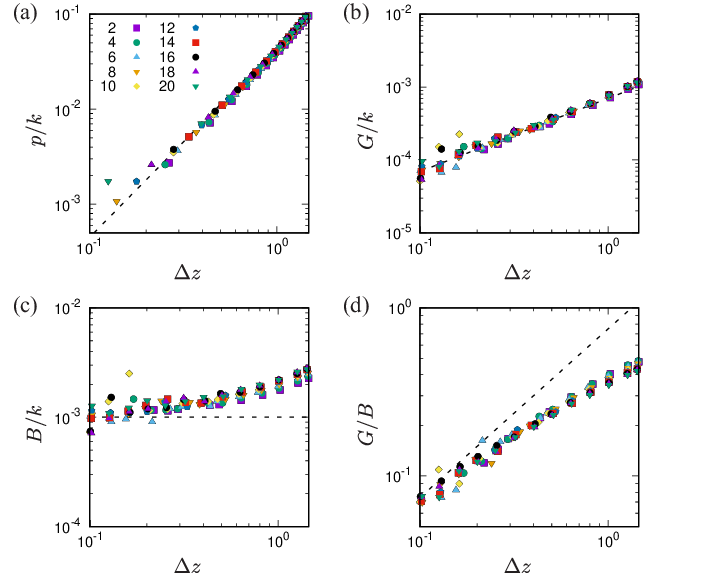}
\caption{
The (a) scaled pressure $p/k$, (b) scaled shear modulus $G/k$, (c) scaled bulk modulus $B/k$, and (d) the ratio $G/B$ as functions of the excess coordination number $\Delta z$.
The size ratio $\lambda$ increases as listed in the legend of (a).
The dashed lines in (a), (b), and (d) represent the critical scaling,\ i.e.\ (a) $p/k\sim\Delta z^2$, (b) $G/k\sim\Delta z$, and (d) $G/B\sim\Delta z$, respectively.
In (c), $B/k$ converges to a constant (dashed line) as the system approaches the unjamming transition, $\Delta z\rightarrow 0$.
\label{fig:mods}}
\end{figure}

\emph{Elastic moduli} ---
Though the dependence of $p/k$ and $\Delta z$ on the packing fraction is influenced by the polydispersity (Fig.\ \ref{fig:last}),
the relation between $p/k$ and $\Delta z$ is independent of the size ratio $\lambda$.
Figure \ref{fig:mods}(a) displays scatter plots of $p/k$ and $\Delta z$, where $\lambda$ increases as listed in the legend.
Strikingly, all the data are nicely collapsed onto the critical scaling, $p/k\sim\Delta z^2$ (dashed line).
In Figs.\ \ref{fig:mods}(b)-(d), we also show elastic moduli,\ i.e.\ shear modulus $G$ and bulk modulus $B$, of polydispersed particles
and their ratio,\ i.e.\ $G/B$, as functions of $\Delta z$. (A discussion of how to calculate the moduli is presented in Appendix \ref{app:matrix}.)
All the data in (b)-(d) are well collapsed.
The scaled shear modulus exhibits the critical scaling, $G/k\sim\Delta z$ (dashed line in (b)).
Moreover, the scaled bulk modulus $B/k$ converges to a constant (dashed line in (c))
and the critical scaling, $G/B\sim\Delta z$ (dashed line in (d)), can be confirmed as the system approaches the unjamming transition,\ i.e.\ as $\Delta z\rightarrow 0$.
Therefore, we conclude that the scaling relations $p/k\sim\Delta z^2$, $G/k\sim\Delta z$, and $G/B\sim\Delta z$, which are all hallmarks of jamming transition, are also {\em insensitive} to polydispersity.
Instead, linear elasticity near jamming is controlled only by the mean coordination number.
This is especially surprising in light of the observation that the coordination distribution $P(z)$ {\em is} sensitive to $\lambda$ (Fig.~\ref{fig:pdfz}).
%

\emph{Normal modes} ---
A system's elastic moduli are determined by its vibrational properties \cite{lemaitre06,rl0}.
We therefore examine whether jammed systems' vibrational properties display the same insensitivity to polydispersity seen in $G$ and $B$.
We calculate eigenfrequencies $\omega$ of polydispersed packings by diagonalizing their dynamical matrix (see Appendix \ref{app:matrix}).
Figure \ref{fig:vdom}(a) displays the vibrational density of states (VDOS, i.e.~distribution function of $\omega$) with the horizontal axis non-dimensionalized by a time unit, $t_0\equiv\sqrt{m_0/k}$ ($m_0$ is the particle mass).
The size ratio is $\lambda=20$, and the excess coordination number increases as indicated by the arrow.
As in the case of bidispersed packings \cite{gn1,gn3,gn4}, the VDOS exhibits a plateau (horizontal dashed line) above a characteristic frequency, $\omega_\ast < \omega$.
As shown in Fig.\ \ref{fig:vdom}(b), the characteristic frequency is linear in the excess coordination number as $\omega_\ast\sim\Delta z$.
In contrast, the VDOS is unaffected by the polydispersity (Fig.\ \ref{fig:vdom}(c)).
The vibrational properties of polydispersed packings are therefore governed only by the coordination number -- just like their elastic moduli.
In Appendix \ref{sec:packing}, we show our results of the participation ratio $P_r(\omega)$,
where $P_r(\omega)$ in intermediate frequencies (the plateau regime in the VDOS) slightly decreases with the increase of $\lambda$.
%
\begin{figure}
\includegraphics[width=\columnwidth]{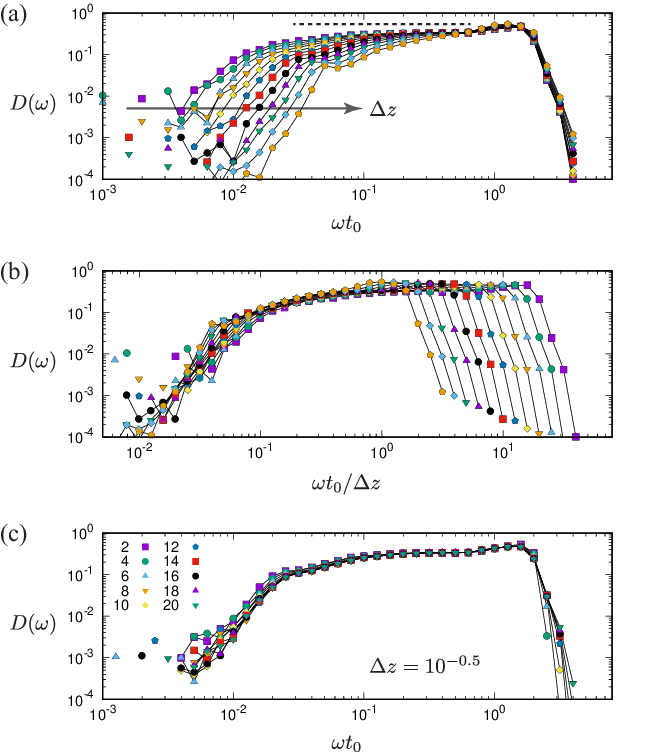}
\caption{
(a) The VDOS of polydispersed particle packings, where the size ratio is given by $\lambda=20$.
The excess coordination number increases from $\Delta z = 10^{-1}$ to $10^{0.1}$ (symbols) as indicated by the arrow.
(b) A scaling data collapse of the VDOS, where the horizontal axis is divided by the excess coordination number $\Delta z$.
(c) The VDOS with $\Delta z = 10^{-0.5}$, where $\lambda$ increases as listed in the legend.
\label{fig:vdom}}
\end{figure}

\emph{Summary \& outlook} ---
In this study, we have numerically investigated polydispersed particle packings in two dimensions.
We found that the distributions of contact force broadens with increasing polydispersity $\lambda$.
The contact number distribution also broadens, developing a power law tail with a characteristic cutoff $z^* \sim \lambda^{0.74}$. 
The jamming density $\phi_c$ monotonically increases as $\phi_c-\phi_c^\ast \sim \left(\lambda-\lambda^\ast\right)^{0.32}$.
In contrast, the critical scaling of pressure and elastic moduli 
are unaffected by $\lambda$.
Furthermore, the VDOS is independent of polydispersity and the critical scaling of characteristic frequency, $\omega_\ast\sim\Delta z$, is the same with the results of monodisperse and bidisperse systems.
Therefore, the mechanical properties and normal modes of soft particle packings are not affected by the particle size distribution and governed only by the excess coordination number.

In our MD simulations, we employed a minimal model of polydispersed particles,
where every contact force is proportional to the same stiffness $k$ and every mass $m_0$ is unique.
Because the variance of particle mass,\ i.e.,\ $m_i$ ($i=1,\dots,N$), merely rescales each row of the dynamical matrix,
we do not expect that distributions of particle mass significantly change the VDOS and elastic moduli.
As demonstrated in random elastic networks \cite{emt1}, however, the scaling of VDOS and characteristic frequency $\omega_\ast$ is controlled by the stiffness distribution.
Thus, the effect of stiffness distribution on our results have to be examined, which we leave as a future work.
Moreover, further analysis in three dimensions is useful for practical applications of this work
and linear viscoelastic properties of polydispersed particles are also an interesting topic for future works.
%
\begin{acknowledgments}
We thank K. Yoshii, H. Katsuragi, and H. Ikeda for fruitful discussions.
This work was financially supported by JSPS KAKENHI Grant Number 22K03459 and the Information Center of Particle Technology.
\end{acknowledgments}
\appendix
\section{Additional results of numerical simulations and normal mode analysis}
\label{sec:packing}
In this appendix, we show additional results of our molecular dynamics (MD) simulations of polydispersed soft particles.
We explain size distributions of the particles (Sec.\ \ref{sub:size})
and examine how elastic energy and mean overlap depend on the polydispersity and packing fraction of the particles (Sec.\ \ref{sub:energy}).
We also analyze static structures of polydispersed soft particles (Sec.\ \ref{sub:struct}) and show our results of participation ratio (Sec.\ \ref{sub:normal}).
%
\subsection{Size distributions}
\label{sub:size}
\begin{figure}
\includegraphics[width=\columnwidth]{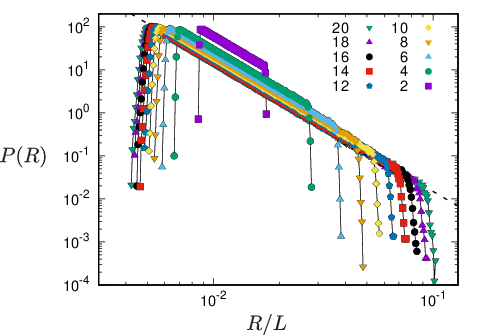}
\caption{
Distribution functions of particle radius $R$, where the horizontal axis is scaled by the system length $L$.
The size ratio, $\lambda=R_\mathrm{max}/R_\mathrm{min}$, decreases as listed in the legend.
The system size (the number of particles) is $N=2048$ and the packing fraction of the particles is given by $\phi=0.90$.
The dashed line indicates the power-law,\ Eq.\ (\ref{eq:size}).
\label{fig:size}}
\end{figure}

In our MD simulations, we randomly sample each particle radius $R$ from a power-law distribution function,
\begin{equation}
P(R) \propto R^{-\nu}~.
\label{eq:size}
\end{equation}
The power-law exponent is given by $\nu=3$
and the distribution function is defined in the range, $R_\mathrm{min}<R<R_\mathrm{max}$.
Figure \ref{fig:size} shows the distribution functions $P(R)$, where the particle radius (horizontal axis) is scaled by the system length as $R/L$.
The dashed line indicates the power-law,\ Eq.\ (\ref{eq:size}),
and the symbols represent the size ratio defined as $\lambda\equiv R_\mathrm{max}/R_\mathrm{min}$ (as listed in the legend).
%
\subsection{Elastic energy and mean overlap}
\label{sub:energy}
To make a static packing of polydispersed soft particles, we randomly distribute the $N$ particles in a $L\times L$ square periodic box,
where packing fraction of the particles is given by $\phi=\sum_{i=1}^N\pi R_i^2/L^2$.
We then minimize elastic energy of the system,
\begin{equation}
E=\sum_{i=1}^N\sum_{j>i}\frac{k}{2}\delta_{ij}^2~,
\label{eq:E}
\end{equation}
by the FIRE algorithm, where $k$ is the stiffness for contact forces and $\delta_{ij}>0$ is an overlap between the particles, $i$ and $j$, in contact.
We stop the energy minimization if every magnitude of the force, $|\bm{f}_i|$ ($i=1,\dots,N$), becomes lower than a threshold,\ i.e.\
\begin{equation}
|\bm{f}_i|<10^{-9}kL~\hspace{1mm}~\mathrm{for}~\hspace{1mm}~\forall i~.
\label{eq:threshold}
\end{equation}

\begin{figure}
\includegraphics[width=\columnwidth]{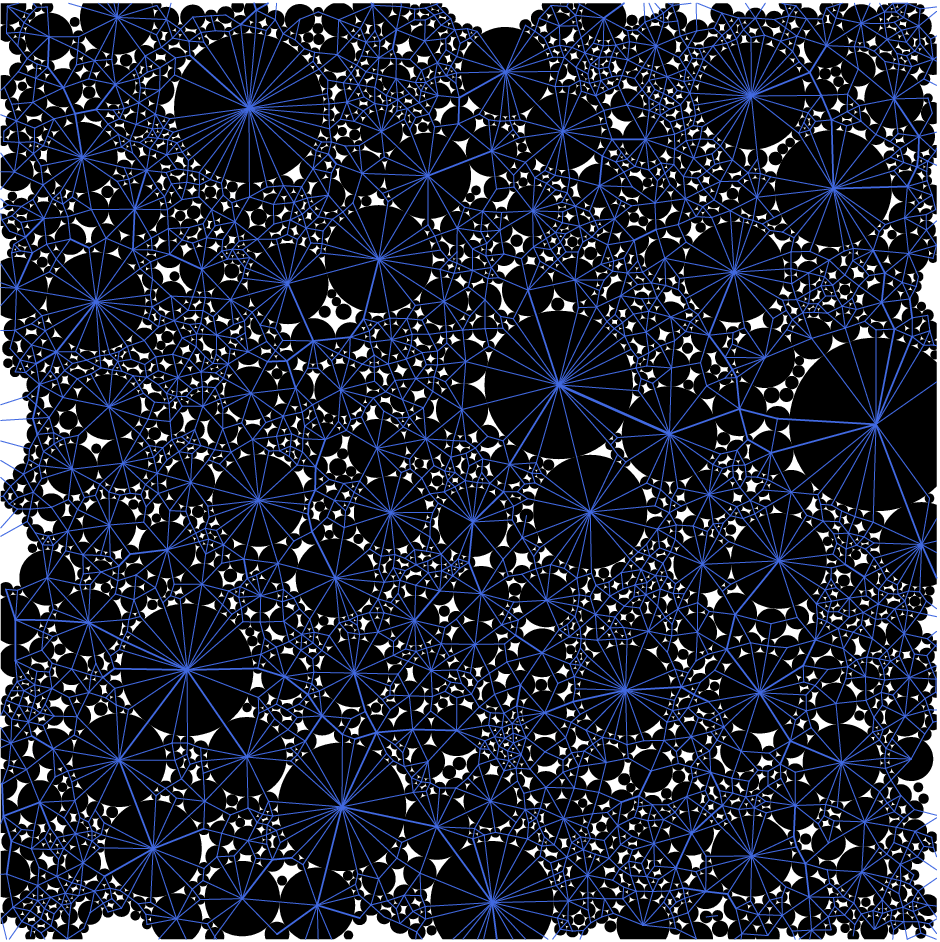}
\caption{
A full image of a static packing of polydispersed soft particles, where the size ratio is $\lambda=20$ and the packing fraction is $\phi=0.90$.
The solid lines represent force-chains, where their width is proportional to the magnitude of repulsive force between the particles in contact.
\label{fig:fchain_full}}
\end{figure}

\begin{figure}
\includegraphics[width=\columnwidth]{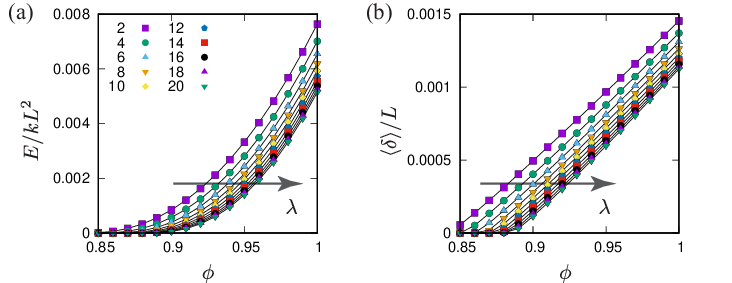}
\caption{
(a) The scaled elastic energy $E/kL^2$ and (b) scaled mean overlap $\langle\delta\rangle/L$ as functions of the packing fraction $\phi$.
The size ratio $\lambda$ increases as listed in the legend of (a) and indicated by the arrows.
\label{fig:last_pot_mdl}}
\end{figure}

Figure \ref{fig:fchain_full} displays a full image of a static packing of $N=2048$ particles (after the energy minimization), where we used $\lambda=20$ and $\phi=0.90$.
The solid lines represent a force-chain network, where their width is proportional to the magnitude of repulsive force,\ i.e.\ $k\delta_{ij}$.
Figure \ref{fig:last_pot_mdl} plots (a) the minimized elastic energy and (b) mean value of the overlaps $\delta_{ij}$ (after the energy minimization) as functions of $\phi$,
where we took ensemble averages of $E$ and $\langle\delta\rangle$ (for each $\phi$) over $1000$ different configurations of the particles.
The size ratio $\lambda$ increases as listed in the legend of (a).
As can be seen, both the elastic energy and mean overlap increase from zero at the jamming transition density, $\phi=\phi_c$,
where $\phi_c$ shifts to higher values as the polydispersity (size ratio $\lambda$) increases.
%
\subsection{Static structures}
\label{sub:struct}
\begin{figure}
\includegraphics[width=\columnwidth]{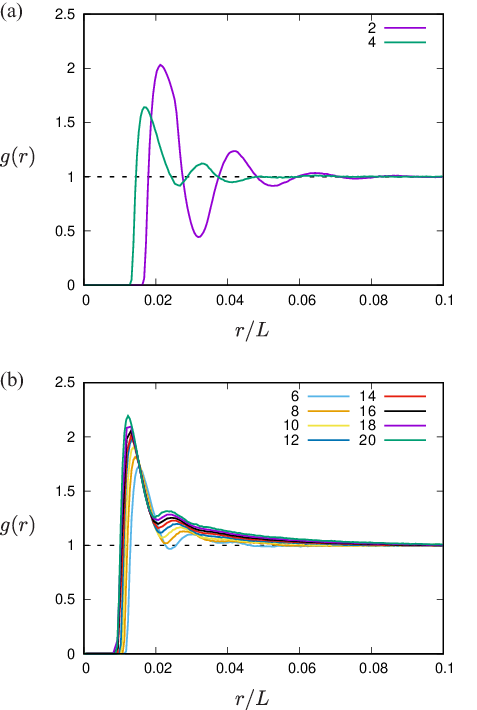}
\caption{
Radial distribution functions of polydispersed soft particles, where the distance $r$ is scaled by the system length as $r/L$ (horizontal axis).
The size ratio $\lambda$ increases as listed in the legends.
\label{fig:mpgr}}
\end{figure}

We examine the influence of polydispersity on static structures of polydispersed soft particles.
Figure \ref{fig:mpgr} shows radial distribution functions $g(r)$ of the particles, where $\lambda$ increases as listed in the legends.
In our simulations, the majority of polydisperse particles are small particles.
Thus, the first peak of $g(r)$ mainly represents the contacts between two small particles.
Because we fix the system length $L$ and control the packing fraction $\phi$,
the smallest particle radius $R_\mathrm{min}$ decreases (and the largest particle radius $R_\mathrm{max}$ increases) with the increase of the size ratio $\lambda=R_\mathrm{max}/R_\mathrm{min}$.
Therefore, as shown in Fig.\ \ref{fig:mpgr}(b), both the first and second peaks shift to shorter distances with the increase of $\lambda$.
In addition, if $\lambda$ is sufficiently large, the second peak of $g(r)$ almost merges with the first peak and
\[
g(r)>1
\]
if $r$ is larger than the distance at the first peak (Fig.\ \ref{fig:mpgr}(b)).
This means that the variation in the distance between two particles is continuous as expected in highly polydisperse systems.

In Fig.\ \ref{fig:pdfz}, we show all the data sets of (a) force distribution $P(f)$ and (b) distribution function of coordination number, $P(z)$.
The force distribution with small polydispersity,\ e.g.\ with $\lambda=2$, is well fitted to a Gaussian distribution (solid line in (a)),
while one observes an exponential tail (dashed line in (a)) if the system is highly polydispersed,\ e.g. if $\lambda=20$.
The distribution function of coordination number broadens with the polydispersity, where the data of $\lambda=20$ exhibits the power-law decay, $P(z)\sim z^{-4.2}$ (dashed line in (b)).

\begin{figure}
\includegraphics[width=\columnwidth]{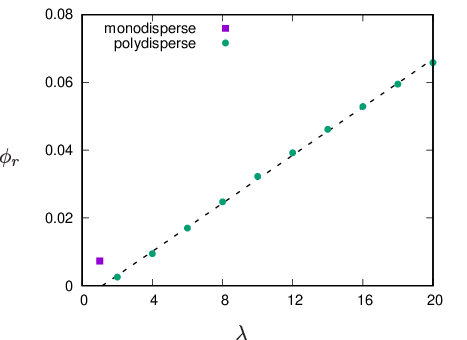}
\caption{
The fraction of \emph{rattlers} $\phi_r$ as a function of the size ratio $\lambda$, where the packing fraction of the particles is fixed to $\phi=0.90$.
The dashed line indicates a linear increase of $\phi_r$.
\label{fig:frr_lambda}}
\end{figure}

We also calculate fraction of \emph{rattlers} as $\phi_r$.
Figure \ref{fig:frr_lambda} displays $\phi_r$ as a function of the size ratio $\lambda$, where the packing fraction of the particles is fixed to $\phi=0.90$.
As can be seen, $\phi_r$ linearly increases with $\lambda$ (except for the monodisperse system, $\lambda=1$).
%
\begin{figure}
\includegraphics[width=\columnwidth]{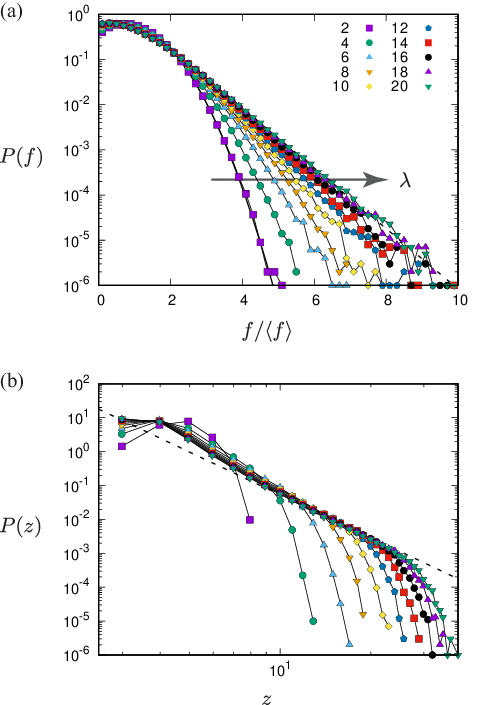}
\caption{
(a) Semi-logarithmic plots of the distribution function of contact force $f$ and (b) double logarithmic plots of the distribution function of coordination number $z$.
The packing fraction of the particles is given by $\phi=0.90$ and the size ratio $\lambda$ increases as indicated by the arrow in (a) and listed in the legend of (a).
In (a), $f$ is scaled by the average $\langle f\rangle$ for each $\lambda$ and the solid line is a Gaussian fit to the data of $\lambda=2$.
The dashed lines indicate (a) the exponential tail and (b) power-law decay, $P(z)\sim z^{-4.2}$, for the data of $\lambda=20$.
\label{fig:pdfz}}
\end{figure}
%
\subsection{Participation ratio}
\label{sub:normal}
Introducing the dynamical matrix of the particles,
we compute its eigenvalues and eigenvectors as $\lambda_n$ and $|n\rangle$, respectively, where $n=1,\dots,Nd$ in $d=2$ dimensions.
We decompose $|n\rangle$ as
\[
|n\rangle = \left(\bm{e}_{1,n},\dots,\bm{e}_{N,n}\right)^\mathrm{T}
\]
with the particle displacement, $\bm{e}_{i,n}$ ($i=1,\dots,N$), associated with the $n$-th mode.
Then, we calculate participation ratio as
\begin{equation}
P_r(n) = \frac{\left(\sum_{i=1}^N \bm{e}_{i,n}^2\right)^2}{N\sum_{i=1}^N |\bm{e}_{i,n}|^4}~.
\label{eq:pr}
\end{equation}

\begin{figure}
\includegraphics[width=\columnwidth]{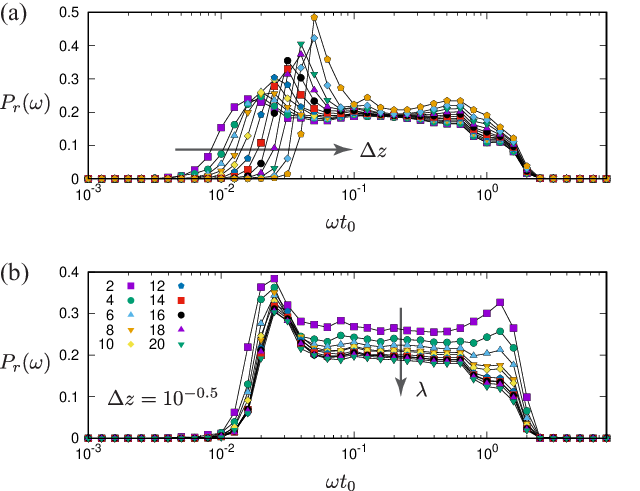}
\caption{
(a) Semi-logarithmic plots of the participation ratio, where the size ratio is given by $\lambda=20$.
The excess coordination number increases from $\Delta z = 10^{-1}$ to $10^{0.1}$ as indicated by the arrow.
(b) The dependence of participation ratio on the size ratio $\lambda$, where the excess coordination number is $\Delta z = 10^{-0.5}$.
The size ratio $\lambda$ increases as listed in the legend and indicated by the arrow.
\label{fig:apma}}
\end{figure}

Figure \ref{fig:apma} displays the participation ratio as a function of eigen-frequency $\omega_n$, where the eigen-frequency is non-dimensionalized by the time unit, $t_0\equiv\sqrt{m_0/k}$.
As can be seen, the participation ratio depends on both (a) the distance from jamming and (b) polydispersity.
%
\section{Dynamical matrix and elastic moduli}
\label{app:matrix}
In this appendix, we introduce \emph{dynamical matrix} (Hessian) and derive elastic moduli from its eigenvalues and eigenvectors.
First, we show explicit forms of the dynamical matrix (Sec.\ \ref{app:sub:matrix}).
Next, we introduce a constitutive relation (Sec.\ \ref{app:sub:constitutive}) and the \emph{extended Hessian} (Sec.\ \ref{app:sub:overdamp}) for isotropic (de)compression.
Then, we formulate the elastic moduli (Sec.\ \ref{app:sub:nmode}).
%
\subsection{Dynamical matrix}
\label{app:sub:matrix}
The $2N\times2N$ dynamical matrix (Hessian) $\mathcal{H}$ consists of second derivatives of the elastic energy $E$ (Eq.\ (\ref{eq:E}))
with respect to the particle positions, $\bm{r}_i=(x_i,y_i)$, as
\begin{equation}
\mathcal{H} =
\begin{pmatrix}
	\frac{\partial^2E}{\partial x_i\partial x_j} & \frac{\partial^2E}{\partial x_i\partial y_j} \\
	\frac{\partial^2E}{\partial y_i\partial x_j} & \frac{\partial^2E}{\partial y_i\partial y_j}
\end{pmatrix}_{i,j=1,\dots,N}~.
\label{eq:hessian}
\end{equation}
We rewrite the elastic energy (Eq.\ (\ref{eq:E})) as the sum of pairwise potentials as
\begin{equation}
E=\sum_{i=1}^N\sum_{j>1}e_{ij}~,
\label{eq:E_pair}
\end{equation}
where each potential is defined as $e_{ij}\equiv k\delta_{ij}^2/2$~($\delta_{ij}>0$).
The second derivatives of $e_{ij}$ are given by
\begin{eqnarray}
\frac{\partial^2 e_{ij}}{\partial x_i\partial x_i} &=& k\left(1-a_{ij}n_{ijy}^2\right)~, \label{ederiv1} \\
\frac{\partial^2 e_{ij}}{\partial x_i\partial y_i} &=& k a_{ij} n_{ijx}n_{ijy}~, \label{ederiv2} \\
\frac{\partial^2 e_{ij}}{\partial y_i\partial y_i} &=& k\left(1-a_{ij}n_{ijx}^2\right)~, \label{ederiv3}
\end{eqnarray}
where we introduced a factor as
\[
a_{ij}\equiv 1+\frac{\delta_{ij}}{r_{ij}}
\]
and $\bm{n}_{ij}\equiv\bm{r}_{ij}/r_{ij}=(n_{ijx},n_{ijy})$ is a unit vector parallel to the relative position between the particles, $i$ and $j$, in contact.
Note that the second derivatives with different indexes ($i\neq j$) satisfy the following relation,
\begin{equation}
\frac{\partial^2 e_{ij}}{\partial\alpha_i\partial\beta_j} = -\frac{\partial^2 e_{ij}}{\partial\alpha_i\partial\beta_i} \hspace{5mm} (\alpha,\beta=x,y)~.
\label{ederiv4}
\end{equation}
Each element in Eq.\ (\ref{eq:hessian}) is constructed from Eqs.\ (\ref{eq:E_pair})-(\ref{ederiv4}).
%
\subsection{Constitutive relation}
\label{app:sub:constitutive}
If isotropic (de)compression is applied to an elastic body in $d$-dimension, strain tensor is written as
\begin{equation}
\epsilon_{\alpha\beta} = \frac{\epsilon}{d}\delta_{\alpha\beta}~,
\label{eq:strain_tensor}
\end{equation}
where $\epsilon\ll 1$ represents a small strain.
We introduce stress tensor and bulk modulus as $\sigma_{\alpha\beta}$ and $B$, respectively, such that constitutive relation is given by
\begin{equation}
\epsilon_{\alpha\alpha} = \frac{1}{dB}\sigma_{\alpha\alpha}~.
\label{eq:constitutive}
\end{equation}
Here, the Einstein summation convention has been used for the subscript $\alpha$.
The confining pressure is defined as
\[
p \equiv -\frac{1}{d}\sigma_{\alpha\alpha}
\]
so that Eq.\ (\ref{eq:constitutive}) is rewritten as
\begin{equation}
p = -B\epsilon_{\alpha\alpha} = -B\epsilon~.
\label{eq:pressure_strain}
\end{equation}
\subsection{The extended Hessian}
\label{app:sub:overdamp}
We assume that the system consisting of $N$ soft particles in $d=2$ dimensions is initially in mechanical equilibrium.
If we apply the strain,\ Eq.\ (\ref{eq:strain_tensor}), to the system, every particle exhibits affine displacement.
Because force balance is broken by the affine displacements, the particles start to move.
After the system relaxes to a static state, every contact force is balanced again,
where particle displacements are now given by the sum of affine and \emph{non-affine displacements}.
We write the non-affine displacement of the $i$-th particle as $\delta\bm{u}_i$ and introduce a $dN=2N$ dimensional vector as
\[
|\delta\bm{u}\rangle \equiv \left(\delta\bm{u}_1,\dots,\delta\bm{u}_N\right)^\mathrm{T}~.
\]

We combine the force balance equation (after the relaxation) and the constitutive relation as
\begin{eqnarray}
\begin{pmatrix}
	|\bm{0}\rangle \\
	V\Delta p/2 \\
	V\Delta p/2
\end{pmatrix}
&=& -
\begin{pmatrix}
	\mathcal{H} & |\Xi_{xx}\rangle & |\Xi_{yy}\rangle \\
	\langle\Xi_{xx}| & \psi_{xx}^{(1)} & \psi_{yy}^{(1)} \\
	\langle\Xi_{yy}| & \psi_{xx}^{(4)} & \psi_{yy}^{(4)}
\end{pmatrix}
\begin{pmatrix}
	|\delta\bm{u}\rangle \\
	\epsilon/2 \\
	\epsilon/2
\end{pmatrix} \nonumber\\
&\equiv& -\mathcal{H}_\mathrm{ex}
\begin{pmatrix}
	|\delta\bm{u}\rangle \\
	\epsilon/2 \\
	\epsilon/2
\end{pmatrix}~.
\label{eq:force_balance}
\end{eqnarray}
Here, $|\bm{0}\rangle$ is the $2N$ dimensional zero vector and $\mathcal{H}$ is the dynamical matrix (Sec.\ \ref{app:sub:matrix}).
In addition, $\Delta p \equiv p-p_0$ with the initial pressure $p_0$ (before deformation) is the increment of confining pressure.
We also introduced the system volume (area) and \emph{extended Hessian} as $V$ and $\mathcal{H}_\mathrm{ex}$, respectively.
In Eq.\ (\ref{eq:force_balance}), the two elements are identical, $\psi_{yy}^{(1)}=\psi_{xx}^{(4)}$,
so that the extended Hessian $\mathcal{H}_\mathrm{ex}$ is a real symmetric matrix.
%
\subsection{Bulk modulus}
\label{app:sub:nmode}
If we rewrite the $dN+2=2N+2$ dimensional vectors in the force balance equation (\ref{eq:force_balance}) as
\begin{eqnarray}
\begin{pmatrix}
	|\delta\bm{u}\rangle \\
	\epsilon/2 \\
	\epsilon/2
\end{pmatrix}
&\equiv&
\left| q, \epsilon/2, \epsilon/2 \right\rangle~,\nonumber\\
\begin{pmatrix}
	|\bm{0}\rangle \\
	V\Delta p/2 \\
	V\Delta p/2
\end{pmatrix}
&\equiv&
\frac{\Delta p}{2}V \left|0,1,1\right\rangle~,\nonumber
\end{eqnarray}
Eq.\ (\ref{eq:force_balance}) is written as
\begin{equation}
\mathcal{H}_\mathrm{ex}\left|q,\epsilon/2,\epsilon/2 \right\rangle
= -\frac{\Delta p}{2}V \left|0,1,1\right\rangle~.
\label{eq:overdamped_omega}
\end{equation}
In addition, the strain is given by an inner product,
\begin{equation}
\epsilon = \langle 0,1,1\left|q,\epsilon/2,\epsilon/2 \right\rangle~.
\label{eq:strain_inner}
\end{equation}

We expand the vector, $\left| q, \epsilon/2, \epsilon/2 \right\rangle$, into the series as
\begin{equation}
\left| q, \epsilon/2, \epsilon/2 \right\rangle = \sum_{n=1}^{2N+2} a_n|n\rangle~,
\label{eq:expansion}
\end{equation}
where $|n\rangle$ is the eigenvector of the extended Hessian and $a_n$ is introduced as a coefficient.
Then, substituting Eq.\ (\ref{eq:expansion}) into (\ref{eq:overdamped_omega}), we find
\begin{equation}
\sum_{n=1}^{2N+2} a_n \mathcal{H}_\mathrm{ex}|n\rangle = -\frac{\Delta p}{2}V \left|0,1,1\right\rangle~.
\label{eq:expansion_1}
\end{equation}
Multiplying the left eigenvector $\langle n|$ to Eq.\ (\ref{eq:expansion_1}), we find
\begin{equation}
a_n\langle n|\mathcal{H}_\mathrm{ex}|n\rangle = -\frac{\Delta p}{2}V \langle n\left|0,1,1\right\rangle~.
\label{eq:expansion_2}
\end{equation}
Because the eigenfrequency $\omega_n$ is given by
\[
\langle n|\mathcal{H}_\mathrm{ex}|n\rangle = \omega_n^2~,
\]
Eq.\ (\ref{eq:expansion_2}) is rewitten as
\[
a_n \omega_n^2 = -\frac{\Delta p}{2} \Lambda_n~,
\]
where we have introduced $\Lambda_n \equiv V\langle n\left|0,1,1\right\rangle$.
Therefore, the coefficient for the expansion is given by
\begin{equation}
a_n = -\frac{\Lambda_n}{2\omega_n^2}\Delta p~.
\label{eq:expansion_3}
\end{equation}

Substituting Eq.\ (\ref{eq:expansion}) into (\ref{eq:strain_inner}), we find that the strain is given by
\begin{eqnarray}
\epsilon &=& \sum_{n=1}^{2N+2} a_n\langle 0,1,1|n\rangle \nonumber\\
&=& \frac{1}{V}\sum_{n=1}^{2N+2} a_n\Lambda_n^\ast \nonumber\\
&=& -\frac{\Delta p}{2V}\sum_{n=1}^{2N+2} \frac{|\Lambda_n|^2}{\omega_n^2}~, \nonumber
\end{eqnarray}
where we used Eq.\ (\ref{eq:expansion_3}).
Therefore, from the constitutive relation,\ Eq.\ (\ref{eq:pressure_strain}), the inverse of bulk modulus is given by
\begin{eqnarray}
\frac{1}{B} &=& \frac{1}{2V}\sum_{n=1}^{2N+2} \frac{|\Lambda_n|^2}{\omega_n^2} \nonumber\\
&\sim& \int_0^\infty \frac{|\Lambda(\omega)|^2}{\omega^2}D(\omega)d\omega~,
\label{eq:inverse_K}
\end{eqnarray}
where $D(\omega)$ is the vibrational density of states (VDOS).
We use our numerical data of $|\Lambda(\omega)|^2$ and $D(\omega)$ to calculate $B$ from Eq.\ (\ref{eq:inverse_K}).
%
%
\bibliography{vdos}
\end{document}